\documentclass{iopart}
\usepackage{iopams}
\usepackage{graphicx,epsf}
\usepackage{color}
\usepackage{mathrsfs}
\usepackage{dcolumn}
\usepackage{cite}

\usepackage{subfig}

\begin{document}

\title[]{Nonlinear excitation of geodesic acoustic mode by  toroidal Alfv\'en eigenmodes and impact on plasma performance}

\author{Zhiyong Qiu$^{1}$, Liu  Chen$^{1, 2}$, Fulvio Zonca$^{3, 1}$ and Wei Chen$^4$}

\address{$^1$Institute for    Fusion Theory and Simulation and Department of Physics, Zhejiang University, Hangzhou, P.R.C}
\address{$^2$Department of   Physics and Astronomy,  University of California, Irvine CA 92697-4575, U.S.A.}
\address{$^3$ ENEA, Fusion and Nuclear Safety Department, C. R. Frascati, Via E. Fermi 45, 00044 Frascati (Roma), Italy}
\address{$^4$ Southwestern Institute of Physics - P.O. Box 432 Chengdu 610041, P.R.C.}

\begin{abstract}
Spontaneous nonlinear excitation of geodesic acoustic mode (GAM) by toroidal Alfv\'en eigenmode (TAE) is investigated using nonlinear gyrokinetic theory. It is found that, the nonlinear decay process depends on thermal ion $\beta_i$ value. Here, $\beta$ is the plasma thermal to magnetic pressure ratio. In the low-$\beta$ limit, TAE decays into a GAM and a lower TAE sideband in the toroidicity induced shear Alfv\'en wave continuous spectrum gap; while in the high-$\beta_i$ limit, TAE decays into a GAM and a propagating kinetic TAE in the continuum. Both cases are investigated for the spontaneous decay conditions. The nonlinear saturation levels of both GAM and   daughter wave are derived. The  corresponding power balance and wave particle power transfer to thermal plasma are computed. Implications on thermal plasma heating are also discussed.
\end{abstract}

\maketitle

\section{Introduction}\label{sec:intro}

Energetic  particle (EP) related physics is a key concern in burning plasmas of next generation magnetic confinement fusion devices such as ITER \cite{KTomabechiNF1991}, characterized by  plasma self heating due to fusion alpha particles \cite{IPBNF1999,AFasoliNF2007}. Because of their high birth energy, fusion alpha particles   heat more effectively electrons than fuel ions via Coulomb collisions. Meanwhile, thermal ion energy is a key control parameter for maximizing the fusion reactivity and, it would be desirable to control fusion alpha particle power transfer and the branching ratio of electron to ion heating. Effective ways for transferring fusion alpha particle power to fuel ions, i.e., so called ``alpha-channeling", have been proposed and investigated \cite{NFischPRL1992,NFischNF1994}.   On the other hand, fusion alpha particles can drive  symmetry breaking electromagnetic perturbations unstable    \cite{LChenRMP2016,YKolesnichenkoVAE1967,AMikhailovskiiSPJ1975,MRosenbluthPRL1975} via resonant wave-particle interactions. For example, shear Alfv\'en wave (SAW) can be excited as instabilities,  lead  to enhanced anomalous alpha particle transport, degradation of plasma performance \cite{IPBNF1999,AFasoliNF2007}  and, potentially, damage plasma facing components due to the heavy heat load \cite{RDingNF2015}.
Due to the equilibrium magnetic field geometry and plasma nonuniformities, SAW instabilities can be excited as Alfv\'en eigenmodes (AEs) inside the frequency gaps of the SAW continuous spectrum, and/or energetic particle continuum modes (EPMs) \cite{IPBNF1999,AFasoliNF2007,LChenRMP2016}. Among various AEs, the well-known toroidal Alfv\'en eigenmode (TAE)
\cite{CZChengAP1985,LChenVarenna1988,GFuPoFB1989} excited inside the toroidicity induced SAW continuum gap is recognized as one of the most serious concerns for the fluctuation-induced EP transport, with the transport rate closely related to TAE saturation amplitude \cite{LChenJGR1999,LChenSpringer2019,MFalessiPoP2018b}. Thus, understanding the nonlinear dynamics of TAE,  including saturation,  is crucial for understanding  the properties of burning plasmas in future reactors, and was under extensive investigation in the past decades \cite{LChenRMP2016,HBerkPoFB1990a,YTodoPoP1995,JZhuPoP2013,SBriguglioPoP2014,DSpongPoP1994, TSHahmPRL1995}. Nonlinear evolution of Alfv\'enic fluctuations, including TAE, can occur along two ``routes" \cite{LChenPoP2013}; i.e., they can saturate through either nonlinear wave-particle phase space dynamics \cite{HBerkPoFB1990a,YTodoPoP1995,JZhuPoP2013,SBriguglioPoP2014,TWangPoP2018b} and/or nonlinear mode-mode coupling processes \cite{DSpongPoP1994, TSHahmPRL1995,LChenPPCF1998,FZoncaPRL1995,LChenPRL2012,ZQiuEPL2013}, as reviewed in Ref. \citenum{LChenRMP2016}.

Axisymmetric zonal structures (ZS)    related physics, including zonal flow \cite{MRosenbluthPRL1998}, zonal current \cite{DSpongPoP1994} and (EP) phase space zonal structures \cite{FZoncaNJP2015}, is another important topic in confined fusion plasma physics research.     ZS  are generally  recognized as the generators of nonlinear equilibria \cite{LChenRMP2016,MFalessiPoP2018b,LChenNF2007a}, and can be driven nonlinearly by micro-scale drift wave (DW) type turbulences including drift Alfv\'en waves (DAWs), and in turn, scatter DW/DAW into radially short wavelength stable regime \cite{ZLinScience1998,LChenPoP2000}.
The nonlinear excitation of ZS, as an important mode-mode coupling channel  for AEs saturation, are investigated in a few recent publications \cite{LChenPRL2012,ZQiuPoP2016,ZQiuNF2017,HZhangPST2013,YTodoNF2010,YTodoNF2012a,YTodoNF2012b}. Noteworthy is that  geodesic acoustic mode (GAM) \cite{NWinsorPoF1968}, as the finite frequency counterpart of zonal flow, can also be excited by TAE \cite{ZQiuEPL2013,ZQiuPRL2018}, leading to nonlinear TAE saturation. GAM is predominantly an electrostatic mode unique to toroidal plasmas, and exists due to the thermal plasma compressibility. GAM is characterized by an $n/m=0/0$ scalar potential and and $n/m=0/1$ up-down anti-symmetric density perturbation, with $n/m$ denoting the toroidal/poloidal mode numbers when using a standard Fourier decomposition of fluctuation fields.
Nonlinear excitation of GAM by TAE was firstly investigated in Ref. \citenum{ZQiuEPL2013}, where a pump TAE decaying into a GAM and a TAE lower sideband inside the frequency gap was studied using nonlinear gyrokinetic theory. It was found that, for spontaneous decay, the pump TAE should lie within the upper half of the toroidicity induced SAW continuum frequency gap, which is not the usual case. In Ref. \citenum{ZQiuPRL2018}, a new decay channel has been proposed and analyzed, i.e., a pump TAE decay into a GAM and a propagating lower kinetic TAE (LKTAE).  LKTAEs are  eigenmodes in the SAW continuum frequency range, which are discretized by kinetic effects, such as finite ion Larmor radius (FLR)  effects and electron parallel dynamics including dissipation \cite{FZoncaPoP1996,FZoncaPoP2014b,RMettPoFB1992,HBerkPoFB1993,JCandyPoP1994,JCandyPPCF1993}. A series of LKTAEs can co-exist, with a small frequency separation.  Note that, because of the frequency matching constraint,  the processes investigated in Ref. \citenum{ZQiuEPL2013} and \citenum{ZQiuPRL2018}  occur, respectively, as $4q^2\beta_i/\epsilon^2$ is smaller or larger than unity; i.e., as GAM frequency is smaller or larger than the distance between the pump TAE frequency and the lower accumulation point frequency of the toroidicity induced SAW continuum frequency gap. Here, $q$ is the safety factor, $\beta_i$ is the ratio of ion thermal pressure to equilibrium magnetic field pressure, and $\epsilon$ is the inverse aspect ratio. It is found that  the process proposed in Ref. \citenum{ZQiuPRL2018} is relevant and possibly important for typical   burning plasma  parameters, and can influence not only the EP confinement via TAE saturation  but also the nonlocal  power transfer from fusion alpha particles to thermal plasma via  ion Landau damping of the nonlinearly driven GAM \cite{HSugamaJPP2006,ZQiuPPCF2009}. The secondary GAM excited by TAE, being the finite frequency counterpart of zonal flow, may also regulate DW turbulence \cite{FZoncaEPL2008,ZQiuPoP2014}, and cause cross-scale couplings and  confinement improvement \cite{TSHahmPoP1999}.

In this work, using gyrokinetic theory, we present a detailed analysis of nonlinear excitation of GAM by TAE, and discuss the saturation level of AEs and ZS, as well as the corresponding power balance and the power transfer from EPs to thermal plasmas via different channels. Thereby,  we address the impact of TAE decay by GAM on plasma performance including fuel ion heating.
The rest of the paper is organized as follows. In Sec. \ref{sec:model}, the theoretical model is given. The parametric process is investigated in Sec. \ref{sec:nl_dr}. The effect of this process on plasma heating is discussed in Sec. \ref{sec:heating}. And finally, a summary is given in Sec. \ref{sec:summary}.

\section{Theoretical model}\label{sec:model}

We  investigate the nonlinear interactions among   pump TAE ($\mathbf{\Omega}_0\equiv(\omega_0,\mathbf{k}_0)$),   GAM ($\mathbf{\Omega}_G\equiv(\omega_G,\mathbf{k}_G)$) and   high frequency daughter wave ($\mathbf{\Omega}_h\equiv(\omega_h,\mathbf{k}_h)$) \footnote{Here, ``high" indicates that the mode frequency is high with respect  to the other daughter wave, GAM, generated during the decay process. The frequency, however, is lower than that of the pump TAE  for the spontaneous decay \cite{RSagdeevbook1969}.}  with the same poloidal and toroidal mode numbers as the pump TAE. Here, the high frequency daughter wave  can be another  TAE within the toroidicity induced SAW continuum gap, as in the case of Ref. \citenum{ZQiuEPL2013}, or a propagating LKTAE in the SAW continuous frequency spectrum  \cite{ZQiuEPL2013,ZQiuPRL2018}, depending on the respective $\beta_i$ regime for the two processes to take place.    For TAE and the high frequency daughter wave, the scalar potential $\delta \phi$ and parallel component of  vector potential $\delta A_{\parallel}$ are taken as the field variables, since the corresponding parallel magnetic perturbation is negligible \cite{LChenRMP2016}.  Furthermore, $\delta\psi\equiv \omega\delta A_{\parallel}/(ck_{\parallel})$ is taken as an alternative variable for TAE and the high frequency daughter wave, and one recovers the ideal MHD constraint by taking $\delta\psi=\delta\phi$. One then has, $\delta\phi=\delta\phi_0+\delta\phi_G+\delta\phi_h$, with the subscripts $0$, $G$ and $h$ denoting pump TAE, GAM and high frequency daughter wave, respectively.
Without loss of generality, $\mathbf{\Omega}_0=\mathbf{\Omega}_G+\mathbf{\Omega}_h$ is adopted as the frequency/wavenumber matching conditions.
Meanwhile, for TAE and   the high frequency daughter wave with high toroidal mode numbers in burning plasmas \cite{LChenRMP2016}, we adopt the well-known ballooning-mode decomposition \cite{JConnorPRL1978} (see, e.g., Refs. \cite{LChenRMP2016,FZoncaPoP2014b,ZLuPoP2012} for a recent review of mode structure representation in toroidal geometry) in the $(r,\theta,\phi)$ field-aligned flux coordinates
\begin{eqnarray}
\delta\phi_0&=&A_0e^{i(n\phi-m_0\theta-\omega_0t)}\sum_j e^{-ij\theta}\Phi_0(x-j)+c.c.,\nonumber\\
\delta\phi_h&=&A_he^{i(n\phi-m_0\theta-\omega_0t)}e^{-i(\int \hat{k}_Gdr-\omega_Gt)}\nonumber\\
&&\hspace*{6em}\times \sum_je^{-ij\theta}\Phi_h(x-j)+c.c..\nonumber
\end{eqnarray}
Here, $(m=m_0 +j, n)$ are the poloidal and toroidal mode
numbers, $m_0$ is the reference value of $m$,
$nq(r_0) =m_0$,   $x=nq-m_0 \simeq
nq'(r-r_0)$,  $\hat{k}_G$ is the radial envelope due to GAM modulation and  $\hat{k}_G\equiv nq'\theta_{k_G}$  in the  ballooning representation, $\Phi$ is the fine radial structure associated with $k_{\parallel}$ and magnetic shear, and $A$ is the radial envelope.

For the predominantly electro-static
GAM characterized by radially corrugated scalar potential, one has
\begin{eqnarray}
\delta\phi_G&=&A_Ge^{i(\int \hat{k}_Gdr-\omega_Gt)}\sum_j \Phi_G(x-j)+c.c..\nonumber
\end{eqnarray}
Here, $\Phi_G$ is the fine scale structure of GAM due to the weakly ballooning features of both the pump TAE and the high frequency daughter wave \cite{ZQiuNF2017}\footnote{The fine radial  scale   of zonal structure due to the weakly ballooning nature of the pump TAE was not accounted for in Ref. \citenum{ZQiuEPL2013}.}, and the summation over $j$ is the summation over the radial positions where the pump TAE poloidal harmonics are  localized. As a result,
$\mathbf{k}_G=\mathbf{\hat{k}}_G-i\partial_r\ln\Phi_G \hat{\mathbf{e}}_r$,
and one  typically has $|\partial_r \ln\Phi_G|\gg |\hat{k}_G|$.
In the expression for pump TAE, GAM and the high frequency daughter wave, frequency and wavenumber matching conditions are implicitly assumed. This is generally valid. For the high-$\beta_i$ limit where the  high frequency daughter wave is an  LKTAE,   the frequency difference between neighbouring LKTAEs is rather small \cite{GVladRNC1999}, and, thus, the LKTAEs
can be considered as a dense spectrum of eigenmodes. Thus, the frequency mismatch effects on the three wave decay process is generally unimportant, consistent with the dense continuum limit of
LKTAEs. In the low-$\beta_i$ limit, the high frequency daughter wave is a TAE lower sideband with finite  radial envelope wave-number ($\theta_{k_G}$), and the matching condition comes from the finite $\theta_k$ dependence of the  TAE lower sideband frequency.

The governing equations for the resonant three wave interactions, can then be derived from quasi-neutrality condition
\begin{eqnarray}
\frac{n_0e^2}{T_i}\left(1+\frac{T_i}{T_e}\right)\delta\phi_k=\sum_s \left\langle q J_k\delta H_k \right\rangle_s,\label{eq:QN}
\end{eqnarray}
and nonlinear gyrokinetic vorticity equation
\begin{eqnarray}
&&\frac{c^2}{4\pi \omega^2_k}B\frac{\partial}{\partial l}\frac{k^2_{\perp}}{B}\frac{\partial}{\partial l}\delta \psi_k +\frac{e^2}{T_i}\left\langle (1-J^2_ k)F_0\right\rangle\delta\phi_k\nonumber\\
&&-\sum_s\left\langle\frac{q}{\omega_k}J_k\omega_d\delta H_k \right\rangle_s\nonumber\\
&=&-i\frac{c}{B_0\omega_k}\sum_{\mathbf{k}=\mathbf{k}'+\mathbf{k}''} \mathbf{\hat{b}}\cdot\mathbf{k}''\times\mathbf{k}'\left [ \frac{c^2}{4\pi}k''^2_{\perp} \frac{\partial_l\delta\psi_{k'}\partial_l\delta\psi_{k''}}{\omega_{k'}\omega_{k''}} \right.\nonumber\\
&&\left.+ \left\langle e(J_kJ_{k'}-J_{k''})\delta L_{k'}\delta H_{k''}\right\rangle \right].
\label{eq:vorticityequation}
\end{eqnarray}
Here, $J_k\equiv J_0(k_{\perp}\rho)$ with $J_0$ being the Bessel function of zero index accounting for FLR effects, $\rho=v_{\perp}/\Omega_c$ is the Larmor radius, $\Omega_c=eB/(mc)$ is the cyclotron frequency, $F_0$ is the equilibrium particle distribution function,  $\sum_s$ is the summation on different particle species,    $\omega_d=(v^2_{\perp}+2
v^2_{\parallel})/(2 \Omega R_0)\left(k_r\sin\theta+k_{\theta}\cos\theta\right)$ is the magnetic drift frequency,  $l$ is the arc length along the equilibrium magnetic field line, $\delta L_k\equiv\delta\phi_k-k_{\parallel} v_{\parallel}\delta\psi_k/\omega_k$; and other notations are standard.  The dominant  nonlinear terms in the vorticity equation are        Maxwell and Reynolds stresses \footnote{Some subtleties with the interpretation of Maxwell and Reynolds stresses in equation (2) are discussed in Ref. \citenum{LChenRMP2016}.}; formally written on the right hand side of equation (\ref{eq:vorticityequation}).  Furthermore, $\langle\cdots\rangle$ indicates velocity space integration and $\delta H_k$ is the nonadiabatic particle response, which can be derived from nonlinear gyrokinetic equation \cite{EFriemanPoF1982}:
\begin{eqnarray}
&&\left(-i\omega+v_{\parallel}\partial_l+i\omega_d\right)\delta H_k=-i\omega_k\frac{q}{T}F_0J_k\delta L_k \nonumber\\
&&\hspace*{4em}-\frac{c}{B_0}\sum_{\mathbf{k}=\mathbf{k}'+\mathbf{k}''}\mathbf{b}\cdot\mathbf{k''}\times\mathbf{k'}J_{k'}\delta L_{k'}\delta H_{k''}\label{eq:NLGKE}.
\end{eqnarray}
In equation (\ref{eq:NLGKE}), the   free energy associated with pressure gradient is neglected in the formally linear term on the right hand side, assuming the free energy  driving the pump TAE unstable comes from the EP pressure gradient, while the nonlinear mode coupling process studied here is dominated by the thermal plasma contribution. For a discussion on the contribution of resonant EPs on ZS generation by TAE \cite{YTodoNF2010}, which may dominate in the linear growth stage of the pump TAE, interested readers may refer to Refs. \cite{ZQiuPoP2016,ZQiuNF2017}.

\section{Parametric instability dispersion relation}\label{sec:nl_dr}

The particle responses to TAE/LKTAE and GAM can be derived   from equation (\ref{eq:NLGKE}), by  taking a small amplitude expansion  $\delta H_k=\delta H^L_k+\delta H^{NL}_k$, with the superscripts ``L" and ``NL" denoting linear and nonlinear responses, respectively. The leading order linear particle responses
are given below, which are then used to derive the nonlinear particle responses.
For pump TAE and the high frequency daughter wave, with $k_{T,\parallel}\simeq 1/(2qR_0)$ and $|\omega_T|\simeq |V_A/(2qR_0)|$, one has $|k_{\parallel}v_{t,e}|\gg |\omega_T|\gg |k_{\parallel}v_{t,i}|\gg|\omega_{d,e}|, |\omega_{d,i}|$, and the linear particle responses to TAE/LKTAE can be derived as
\begin{eqnarray}
\delta H^L_{T,e}&=&-\frac{e}{T_e}F_0\delta\psi_T+\delta K^L_{T,e},\nonumber\\
\delta H^L_{T,i}&\simeq&\frac{e}{T_i}F_0J_T\delta\psi_T+\delta K^L_{T,i}.\nonumber
\end{eqnarray}
The subscript ``$T$" is used for modes in the TAE frequency range, and the expressions are applicable to both TAE and LKTAE. Meanwhile, $\delta K^L_{T,e}$ and $\delta K^L_{T,i}$ account for kinetic compression effects of the thermal plasma (see, e.g., Ref. \cite{LChenRMP2016}), which are not explicitly given here since they are typically of higher order and describe damping as well as diamagnetic effects that are assumed implicitly. The corresponding EP linear responses are also implicitely accounted for \cite{LChenRMP2016}, without being explicitly given in the present work for the sake of simplicity.  Substituting the leading order linear particle responses into the quasi-neutrality condition, one then has
\begin{eqnarray}
\delta\psi_T=\left(1+\tau-\tau\Gamma_T\right)\delta\phi_T\equiv \sigma_T\delta\phi_T.\nonumber
\end{eqnarray}
Here, $\tau\equiv T_e/T_i$ is the electron to ion temperature ratio, $\Gamma_k\equiv \langle J^2_k F_0/n_0 \rangle$,  $\sigma_T\neq 1$ describes the deviation from ideal MHD condition due to kinetic effects and generation of parallel electric field, which is  important in the high-$\beta_i$ limit, when the high frequency daughter wave is an  LKTAE, while $\sigma_0\simeq 1$ for the  pump TAE as well as the TAE lower sideband in the low-$\beta_i$ limit.  Substituting into linear vorticity equation, one has
\begin{eqnarray}
\mathscr{E}_T\delta\phi_T\equiv \left(1-\Gamma_T- k^2_{T,\parallel}V^2_A\sigma_T b_T/\omega^2_T+\Delta_T \right)\delta\phi_T=0.\nonumber
\end{eqnarray}
Here, $\Delta_T$ accounts for thermal plasma as well as EPs compression effects in toroidal geometry, proportional to particle magnetic drifts. $\Delta_T$ (and, hence, $\mathscr{E}_T$) is generally a linear integro-differential operator and its known exact expression \cite{LChenRMP2016,FZoncaPoP2014b,FZoncaPoP1996} is not needed here for the present scope. Thus, we just indicate it formally and recall that its calculation yields expressions for EP drive as well as the thermal plasma collisionless and collisional damping. Note that, for short wavelengths, the previous equation
is the formal dispersion relation of KAW, and it yields $\omega^2=k^2_{\parallel}V^2_A(1+ C_k k^2_{\perp} \rho^2_i)$, with $C_k\equiv 3/4+\tau(1-i\delta_e)$. In particular, the $\delta_e$ term accounts for trapped electron collisional damping \cite{NGorelenkovPS1992}  but, if properly modeled, also includes  electron Landau damping and is responsible for the electron heating by LKTAE, as we will discuss  in Sec. \ref{sec:heating}. However,  this $\delta_e$ term is not explicitly kept in our  derivation of $\delta H^L_{T,e}$,  which aims at giving the lowest order bulk particle response to be used for the nonlinear derivation.  The eigenmode dispersion relation of TAE/LKTAE can be derived, noting the $V^2_A\propto (1-2\epsilon_0\cos\theta)$ dependence of $V^2_A$ due to toroidicity,
and matching the solutions through the radially fast to slowly varying regions \cite{CZChengAP1985,FZoncaPoP1996,FZoncaPoP2014b}. Here, $\epsilon_0=r/R_0+\Delta'$ with $\Delta'$ being the Shafranov shift.

Linear particle response to GAM can be derived, noting the $\omega_G\sim |v_{t,i}/R_0|\gg |\omega_{tr,i}|, |\omega_d|$ ordering based on $q\gg1$, and that $k_{\parallel,G}=0$.   One then has, to the leading order \cite{ZQiuPPCF2009},
\begin{eqnarray}
\delta H^L_{G,e}&=&-\frac{e}{T_e}F_0\overline{\delta\phi}_G,\nonumber\\
\delta H^L_{G,i}&=&\frac{e}{T_i}F_0J_G\delta\phi_G.\nonumber
\end{eqnarray}
Here, $\omega_{tr}\equiv v_{\parallel}/(qR_0)$ is the transit frequency, and $\overline{(\cdots)}\equiv \int d\theta (\cdots)/(2\pi)$ denotes surface averaging.

\subsection{Nonlinear GAM equation}

The nonlinear GAM  equation in the electrostatic limit can be determined from the nonlinear vorticity equation.  One obtains,  after some tedious but straightforward algebra
\begin{eqnarray}
&&\mathscr{E}_{G^*}\overline{\delta\phi_{G^*}}=i\frac{c}{B\omega_G}k_Gk_{0,\theta}\nonumber\\
&&\hspace*{1.em}\times \left[\Gamma_0-\Gamma_h-(b_h-b_0)\frac{k^2_{\parallel}V^2_A}{\omega_0\omega_h}\sigma_{0^*}\sigma_h\right]\overline{\delta\phi_{0^*}\delta\phi_h}, \label{eq:GAM_local}
\end{eqnarray}
with $k_{0,\theta}=-m_0/r$ being the poloidal mode number of the pump TAE $\Omega_0\equiv(\omega_0, \mathbf{k}_0)$.  The two terms on the right hand side of equation (\ref{eq:GAM_local}) are, respectively, the generalized Reynolds and Maxwell stresses, valid for arbitrary $k_{\perp}\rho_i$.
Furthermore, $\mathscr{E}_G$ is the linear dispersion function of GAM,   defined as \cite{FZoncaEPL2008}
\begin{eqnarray}
\mathscr{E}_G\equiv \left\langle (1-J^2_G)\frac{F_0}{n_0}\right\rangle-\left.\frac{T_i}{n_0e^2}\sum_s\overline{\left\langle \frac{q_s}{\omega}J_G\omega_d\delta H^L_{G}\right\rangle}\right/\overline{\delta\phi}_{G}.\nonumber
\end{eqnarray}

Taking $\Phi_{G^*}\equiv\Phi_{0^*}\Phi_h$ as the fast varying component \cite{ZQiuNF2017} of GAM,  one then has,  the GAM eigenmode dispersion relation from the radially slowly varying component of equation (\ref{eq:GAM_local}):
\begin{eqnarray}
\mathscr{E}_{G^*}A_{G^*}=i\frac{c}{B_0}k_{0,\theta}\frac{1}{\omega_G}\hat{\alpha}_GA_{0^*}A_h, \label{eq:GAM_global}
\end{eqnarray}
with
\begin{eqnarray}
\hat{\alpha}_G&\equiv& \int \Phi_{0^*}\Phi_h k_G\left[\Gamma_0-\Gamma_h-(b_h-b_0)\frac{k^2_{\parallel}V^2_A}{\omega_0\omega_h}\sigma_{0^*}\sigma_h\right] dr\nonumber\\
&&\hspace*{4em} \times\left(\int \Phi_{0^*}\Phi_hdr\right)^{-1}.\nonumber
\end{eqnarray}
$\hat{\alpha}_G$ contains the complex information of breaking of pure Alfv\'enic state \cite{LChenPoP2013} by toroidicity \cite{LChenPRL2012,ZQiuEPL2013} and kinetic effects \cite{FZoncaEPL2015}, as well as mode structure due to equilibrium magnetic geometry.
Equation (\ref{eq:GAM_global}) is valid for arbitrary $k_{\perp}\rho_i$.
In the long wavelength   $|k_{\perp}\rho_i|\ll1$  and low-$\beta_i$ limit, equation (\ref{eq:GAM_global}) recovers equation (8) of Ref. \citenum{ZQiuEPL2013}, where GAM excitation by the beating of the pump TAE and a TAE lower sideband  within the toroidicity induced SAW continuum gap is investigated. On the other hand, in the high-$\beta_i$ limit with consequently $k_{\perp}\rho_i\sim O(1)$, equation (\ref{eq:GAM_local}) recovers equation (2) of Ref. \citenum{ZQiuPRL2018}, where GAM was excited by the beating of pump TAE to an LKTAE.

\subsection{Nonlinear high frequency daughter wave equation}

Nonlinear electron response to the high frequency daughter wave can be derived noting the $|k_{\parallel}v_{t,e}|\gg|\omega_T|\gg|\omega_{d,e}|$ ordering. The  gyrokinetic equation for nonlinear electron response to the high frequency daughter wave,  to the leading order, is
\begin{eqnarray}
v_{\parallel}\partial_l\delta H^{NL}_{h,e}&\simeq&-\frac{c}{B_0}\mathbf{b}\cdot\mathbf{k}_0\times\mathbf{k}_{G^*}\left(\delta L_{G^*}\delta H^L_0-\delta L_0\delta H^L_{G^*}\right)\nonumber,
\end{eqnarray}
which can be solved and yields
\begin{eqnarray}
\delta H^{NL}_{h,e}=-i\frac{c}{B_0}\frac{e}{T_e}F_0k_{0,\theta}k_G\frac{1}{\omega_0}\delta\psi_{0}\overline{\delta\phi}_{G^*}.\label{eq:NL_electron_KTAE}
\end{eqnarray}

Nonlinear ion response to the high frequency daughter wave can be derived   noting the $\omega_T\gg k_{\parallel}v_{t,i},\omega_{d,i}$ ordering, and is given as
\begin{eqnarray}
\delta H^{NL}_{h,i}=-i\frac{c}{B_0}k_Gk_{0,\theta}\frac{e}{T_i}F_0 \frac{k_{\parallel,0}v_{\parallel}}{\omega_0\omega_h}J_0J_G\delta\phi_{0}\overline{\delta\phi}_{G^*}.\label{eq:NL_ion_KTAE}
\end{eqnarray}

Substituting equations (\ref{eq:NL_electron_KTAE}) and (\ref{eq:NL_ion_KTAE}) into quasi-neutrality condition, we then have
\begin{eqnarray}
\delta\psi_h=\sigma_h\delta\phi_h-i\frac{c}{B_0}\frac{k_Gk_{0,\theta}}{\omega_0}\overline{\delta\phi}_{G^*}\delta\phi_{0}.\label{eq:QN_KTAE}
\end{eqnarray}
The nonlinear ion response $\delta H^{NL}_{h,i}$ is an odd function of $v_{\parallel}$, and it has no contribution to the quasi-neutrality condition to the leading order. Thus, equation (\ref{eq:QN_KTAE}) describes the nonlinear electron correction to the ideal MHD condition of the high frequency daughter wave, in addition to the linear kinetic  corrections contained in $\sigma_h$.

The nonlinear vorticity equation of the high frequency daughter wave then yields
\begin{eqnarray}
&& \frac{c^2}{4\pi\omega^2}B\frac{\partial}{\partial l}\frac{k^2_{\perp}}{B}\frac{\partial}{\partial l}\delta\psi_h +\frac{n_0e^2}{T_i}(1-\Gamma_h+\Delta_h)\delta\phi_h\nonumber\\
&=& i\frac{c}{B}k_Gk_{0,\theta}\frac{n_0e^2}{T_i}\frac{\Gamma_0-\Gamma_G}{\omega_h}\delta\phi_{0}\delta\phi_{G^*}.\label{eq:vorticity_KTAE}
\end{eqnarray}
In equation (\ref{eq:vorticity_KTAE}),  $\Delta_h$ is the integro-differential operator $\Delta_T$ introduced above, specialized to the high frequency daughter wave. Meanwhile, the   SAW continuum up-shift due to kinetic thermal ion compression is neglected, consistent with  the $\beta\ll1$ ordering.
Substituting equation (\ref{eq:QN_KTAE}) into (\ref{eq:vorticity_KTAE}), we then obtain
\begin{eqnarray}
\mathscr{E}_h\delta\phi_h=i\frac{c}{B}k_{G}k_{0,\theta}\left(\frac{\Gamma_0-\Gamma_G}{\omega_h}{\color{red}-}\frac{1-\Gamma_h}{\sigma_h\omega_0}\sigma_0\right) \delta\phi_{G^*}\delta\phi_{0}.\label{eq:LKTAE}
\end{eqnarray}
Here, $\mathscr{E}_h$ is the wave operator of the high frequency daughter wave, and $\sigma_h=1+\tau-\tau\Gamma_h$.
Noting that $\omega_h=\omega_0-\omega_G$, the nonlinear coupling coefficient of equation (\ref{eq:LKTAE})  recovers  that of equation (10)   of Ref. \citenum{FZoncaEPL2015} for KAW lower sideband  generation by a pump KAW beating with a finite frequency convective cell. More precisely, assuming only the lower sideband generation,    only the electro-static convective cell generation should be considered,   assuming $|\omega_G|\ll|\omega_0|$ in the small $\beta$ limit.

The nonlinear radial envelope equation of the high frequency daughter wave, on the other hand, can be derived by multiplying both sides of equation (\ref{eq:LKTAE}) by $\Phi^*_h$, and integrating over meso- radial scales. One obtains,
\begin{eqnarray}
\hat{\mathscr{E}}_hA_h=i\frac{c}{B_0}k_{0,\theta}\hat{\alpha}_hA_{G^*}A_0,\label{eq:KTAE_global}
\end{eqnarray}
with $\hat{\mathscr{E}}_h$ being the eigenmode dispersion function of the high frequency daughter wave
\begin{eqnarray}
\hat{\mathscr{E}}_h\equiv \int dr |\Phi_h|^2  \mathscr{E}_h,\nonumber
\end{eqnarray}
and
\begin{eqnarray}
\hat{\alpha}_h\equiv \int dr |\Phi_0|^2|\Phi_h|^2 k_G \left(\frac{\Gamma_0-\Gamma_G}{\omega_h}-\frac{1-\Gamma_h}{\sigma_h\omega_0}\sigma_0\right).\nonumber
\end{eqnarray}

Note that, equation (\ref{eq:KTAE_global}) is valid for both  low-$\beta_i$ and high-$\beta_i$  cases. In the low-$\beta_i$ case, equation (\ref{eq:KTAE_global}) recovers the TAE lower sideband nonlinear dispersion relation, i.e., equation (14) of Ref. \citenum{ZQiuEPL2013}, including the effects of GAM fine radial structure \cite{ZQiuNF2017}, which are neglected in Ref. \citenum{ZQiuEPL2013}; while, in the high-$\beta_i$ limit, equation (\ref{eq:KTAE_global}) recovers the LKTAE nonlinear dispersion relation, i.e., equation (4) of Ref. \citenum{ZQiuPRL2018} \footnote{Note that, in Ref. \citenum{ZQiuPRL2018}, there is a typo in the definition of $\hat{\alpha}_L$ (Subscript ``L" is used as in Ref. \cite{ZQiuPRL2018}, since  the high frequency daughter wave is a  lower kinetic TAE), where the second term in the bracket should a negative rather than a positive sign (i.e., $(\Gamma_0-\Gamma_G)/\omega_L+(1-\Gamma_L)\sigma_0/(\sigma_L\omega_0)$).  The final results are not changed by the sign mistake.}.

\subsection{Nonlinear parametric dispersion relation}
The nonlinear dispersion relation can   be derived from equations (\ref{eq:GAM_global}) and (\ref{eq:KTAE_global}) as
\begin{eqnarray}
\hat{\mathscr{E}}_h \mathscr{E}_{G^*}=-\left(\frac{c}{B_0}k_{0,\theta}\right)^2\frac{\hat{\alpha}_G\hat{\alpha}_h}{\omega_G} |A_0|^2. \label{eq:NL_DR_general}
\end{eqnarray}
In the low-$\beta_i$ limit, by taking long wavelength $k_{\perp}\rho_i\ll1$ expansion,  the nonlinear term on the right hand side of equation (\ref{eq:NL_DR_general}) recovers  that of equation (17) of Ref. \citenum{ZQiuEPL2013}, where a pump TAE decaying into a GAM and a TAE lower sideband is discussed, with  the enhanced coupling due to fine radial scale structure of GAM.  Meanwhile,  in the high-$\beta_i$ limit, the lower sideband is an LKTAE in the   SAW continuous frequency spectrum, and equation (\ref{eq:NL_DR_general}) recovers equation (5) of Ref. \citenum{ZQiuPRL2018}. Noting that, in both low- and high-$\beta_i$ limits, the high frequency daughter wave can be considered as a normal mode of the system, one can than expand $\mathscr{E}_G$ and $\hat{\mathscr{E}}_{h}$ along the characteristics of GAM and $\Omega_{h}$. In the local limit, one can write:
\begin{eqnarray}
\mathscr{E}_{G^*}&\simeq&i\partial_{\omega_G}\mathscr{E}_{G,R}(\gamma+\gamma_G)\nonumber\\
&\simeq&-2i b_G \left(\gamma+\gamma_G\right)/\omega_G,\nonumber\\
\hat{\mathscr{E}}_{h}&\simeq& i\partial_{\omega_{h}}\hat{\mathscr{E}}_{h,R}(\gamma+\gamma_{h}),\nonumber
\end{eqnarray}
with $\gamma_G\equiv\mathscr{E}_{G,I}/(\partial_{\omega_G}\mathscr{E}_{G,R})$ being the collisionless damping rate of GAM \cite{HSugamaJPP2006,ZQiuPPCF2009}, and $\gamma_h\equiv\mathscr{E}_{h,I}/(\partial_{\omega_h}\mathscr{E}_{h,R})$ being the dissipation of the lower sideband. Here, the subscripts ``R" and ``I"  denote real and imaginary parts, respectively. The validity of the frequency and wavenumber matching conditions   used here to have simultaneously   $\mathscr{E}_G(\omega_G,k_{r,h})=0$ and  $\hat{\mathscr{E}}_{h,R}(\omega_0-\omega_G,k_{r,h})=0$, are discussed in Sec. \ref{sec:model}.

The parametric dispersion relation can be written as
\begin{eqnarray}
(\gamma+\gamma_G)(\gamma+\gamma_h)=-\left(\frac{c}{B_0}k_{0,\theta}\right)^2\frac{\hat{\alpha}_G\hat{\alpha}_h|A_0|^2} {2b_G\partial_{\omega_h}\hat{\mathscr{E}}_{h,R}},\label{eq:parametric_DR}
\end{eqnarray}
which yields the condition for the spontaneous excitation of the parametric decay instability  from $\gamma=0$,
\begin{eqnarray}
-\left(\frac{c}{B_0}k_{0,\theta}\right)^2\frac{\hat{\alpha}_G\hat{\alpha}_h|A_0|^2} {2b_G\partial_{\omega_h}\hat{\mathscr{E}}_{h,R}}>\gamma_h\gamma_G.\label{eq:threshold}
\end{eqnarray}
Equation (\ref{eq:threshold}) describes TAE spontaneous decay as the nonlinear drive overcomes the dissipation due to GAM and high frequency daughter wave damping, and can be solved for the spontaneous decay condition separately for the low- and high-$\beta_i$ cases.

\subsubsection{Low-$\beta_i$ limit: TAE decay into GAM and TAE lower sideband}\label{sec:low_beta}

\begin{figure}
\includegraphics[width=3.0in]{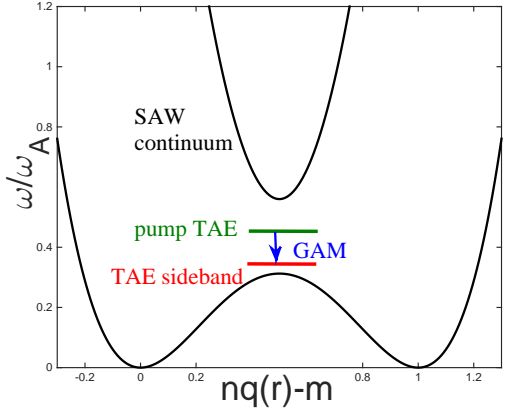}
\caption{TAE decay into GAM and  TAE lower sideband in the low-$\beta_i$ limit. } \label{Fig:low_beta}
\end{figure}

We start from the low-$\beta_i$ limit investigated in Ref. \citenum{ZQiuEPL2013}, where the pump TAE decays into a GAM and a TAE lower sideband, as shown in Fig. \ref{Fig:low_beta}.  Denoting the TAE lower sideband with subscript ``S" and noting that the TAE lower sideband dispersion relation is given as \cite{LChenPRL2012}
\begin{eqnarray}
\hat{\mathscr{E}}_{S}&\equiv& \left(\frac{\omega^4_A}{\epsilon_0\omega^2}\Lambda_T(\omega)D(\omega,k_G)\right)_{\omega=\omega_S},
\end{eqnarray}
with $D(\omega,k_G) = \Lambda_T(\omega)-\delta\hat{W}_f(\omega,k_G)$, $\Lambda_T(\omega)\equiv \sqrt{-\Gamma_-\Gamma_+}$,  $\Gamma_{\pm}\equiv \omega^2/\omega^2_A\pm\epsilon_0\omega^2/\omega^2_A-1/4$ determining the lower and upper accumulational points of toroidicity induced gap,  $\omega^2_A\equiv V^2_A/(q^2R^2_0)$ and $\delta\hat{W}_f(\omega,k_G)$ playing the role of a normalized potential energy \cite{FZoncaPoFB1993}.

Since both TAE and   TAE lower sideband are TAEs with $k_r\rho_i\ll1$, equation (\ref{eq:threshold}) can be greatly simplified by taking $\sigma_0=\sigma_S=1$, and thus,
\begin{eqnarray}
\hat{\alpha}_S &\simeq& -  \frac{2 b_0 k_G}{\omega_0},\nonumber\\
\hat{\alpha}_G &\simeq&  \frac{1}{2}k^3_G\rho^2_i\left(1- \frac{\omega^2_A}{4\omega^2_0}\right),\nonumber
\end{eqnarray}
with the two terms on the right hand side of $\hat{\alpha}_G$ denoting the competition between  Reynolds and Maxwell stresses \cite{LChenPRL2012,LChenPoP2013}. Equation (\ref{eq:parametric_DR})   recovers equation (20) of Ref. \citenum{ZQiuEPL2013}.  Equation (\ref{eq:threshold}) thus, becomes,
\begin{eqnarray}
\gamma_S\gamma_G< &&\left(\frac{c}{B}k_{0,\theta}k_G\right)^2\frac{k^2_{0,\perp}}{k^2_{S,\perp}}\frac{\epsilon_0\omega^3_0}{\omega^4_A\Lambda_T(\omega)}\nonumber\\
&&\times\frac{|A_0|^2}{\partial D/\partial\omega_0}\left( 1-\frac{\omega^2_A}{4\omega^2_0}\right).\label{eq:threshold_low_beta}
\end{eqnarray}
We generally have  $\omega_0\partial_{\omega_0}D(\omega_0,k_G)>0$ in the ideal MHD first stability region for ideal ballooning modes \cite{FZoncaPoFB1993}  and, thus, for the spontaneous decay of TAE into GAM and TAE lower sideband, one requires, first,
\begin{eqnarray}
\omega^2_0>\frac{\omega^2_A}{4},
\end{eqnarray}
i.e., the pump TAE lies within the upper half of the toroidicity induced SAW continuum gap for the nonlinear drive on the right hand side of equation (\ref{eq:threshold_low_beta}) to be positive. Second, the nonlinear drive from pump TAE overcomes the threshold due to GAM and TAE lower sideband damping,
which yields the threshold condition in terms of the pump TAE magnetic perturbation
\begin{eqnarray}
\left(\frac{\delta B_r}{B_0}\right)^2_{thr}\simeq \frac{\gamma_S\gamma_G}{\epsilon_0\omega^2_0}\frac{k^2_{S,\perp}}{k^2_{0,\perp}}\frac{1}{4q^2R^2_0k^2_G}\simeq 10^{-9}-10^{-8}.
\end{eqnarray}
In deriving the above threshold condition, $1-\omega^2_A/(4\omega^2_0)\sim \epsilon_0$ is assumed, while other typical tokamak parameters are used, e.g., $\gamma_S/\omega_0\sim\gamma_G/\omega_G\sim 10^{-2}$, $k_G\rho_i\lesssim 1$, and $k_{\parallel}\rho_i\sim 10^{-3}$ \cite{ZQiuEPL2013}.

\subsubsection{High-$\beta_i$ limit: TAE decay into GAM and LKTAE}\label{sec:high_beta}

\begin{figure}
\includegraphics[width=3.0in]{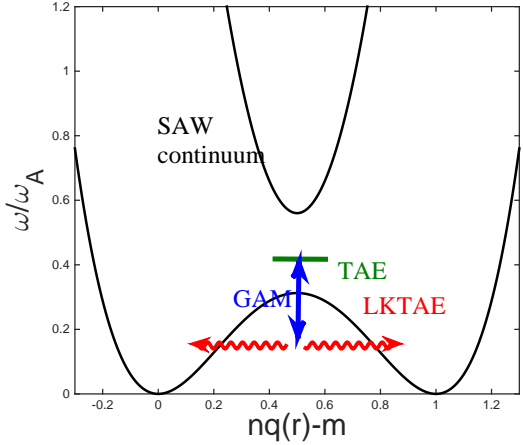}
\caption{TAE decay into GAM and LKTAE in the high-$\beta$ limit.} \label{Fig:high_beta}
\end{figure}

In the high-$\beta_i$ limit, TAE decay into a GAM and a propagating LKTAE in the SAW continuous frequency spectrum, as shown in Fig. \ref{Fig:high_beta}. In the following, we denote the LKTAE with subscript ``L".
For general parameter regimes, especially $k_{\perp}\rho_i\sim 1$ for LKTAE, equation (\ref{eq:threshold}) is an integro-differential equation due to its complex dependence on the mode structure and   equilibrium geometry, and, thus, it requires numerical solution.  However, analytical progress can be made by assuming $b_L\ll 1$.  Noting that, for $|b_k|\ll1$,  $\Gamma_k(b_k)\simeq 1-b_k+3 b^2_k/4$ and $\sigma_k\simeq 1+\tau (b_k+3 b^2_k/4)$,  one then has
\begin{eqnarray}
\hat{\alpha}_G&\simeq&  k_G (b_L-b_0)\left(1-\frac{\omega^2_A}{4\omega_0\omega_L}\right)<0,\nonumber\\
\hat{\alpha}_L&\simeq& \frac{k_G}{\omega_L}\left(b_G-b_0-\frac{b_L}{1+\tau b_L} \frac{\omega_0-\omega_G}{\omega_0}(1+\tau b_0)\right)>0.\nonumber
\end{eqnarray}
Here, $\hat{\alpha}_L$ is positive can be verified noting   that $|k_{r,0}|\sim O(nq'/\epsilon_0)$,  $|k_{r,L}|\simeq O( (\epsilon_0\rho^2_i/(n^2q'^2))^{-1/4})\gg |k_{r,0}|$,  and  $|k_{G}|= |k_{r,0}+k_{r,L}|\simeq|k_{r,L}|$.    Furthermore,
for LKTAE with even mode structure,  the eigenmode dispersion function $\hat{\mathscr{E}}_L$, can be written as \cite{FZoncaPoP1996,FZoncaPoP2014b}
\begin{eqnarray}
\hat{\mathscr{E}}_L\equiv \frac{\pi k^2_{\theta}\rho^2_i\omega^2_A}{2^{2\hat{\xi}+1}\Gamma^2(\hat{\xi}+1/2)\omega^2_L}\left[-\frac{2\sqrt{2}\Gamma(\hat{\xi}+1/2)}{\hat{\alpha} \Gamma(\hat{\xi})}-\delta \hat{W}_f\right],\nonumber\nonumber
\end{eqnarray}
with $\Gamma(\hat{\xi})$ and $\Gamma(\hat{\xi}+1/2)$ being gamma-functions, $\hat{\xi}\equiv 1/4-\Gamma_+\Gamma_-/(4\sqrt{\Gamma_-\hat{s}^2\hat{\rho}^2_K})$, $\hat{\alpha}^2=1/(2\sqrt{\Gamma_-\hat{s}^2\hat{\rho}^2_K})$ with $\hat{\rho}^2_K\equiv (k^2_{\theta}\rho^2_i/2)[3/4+T_e/T_i(1-i\delta_e)]$ denoting the kinetic effects associated with finite ion Larmor radii and electron parallel dynamics including electron dissipation described by $\delta_e$.
One can estimate that $\partial_{\omega_L}\hat{\mathscr{E}}_{L,R}>0$. Thus,  the right hand side of equation (\ref{eq:parametric_DR}) is positive, i.e., pump TAE drives GAM and LKTAE sidebands.

Noting that $|k_{r,0}|\sim O(nq'/\epsilon_0)$,  $|k_{r,L}|\simeq O( (\epsilon_0\rho^2_i/(n^2q'^2))^{-1/4})\gg |k_{r,0}|$,   GAM wave number     $|k_{G}|= |k_{r,0}+k_{r,L}|\simeq|k_{r,L}|$ from matching condition,  and that  $|\delta B_r|\simeq |k_{\theta}\delta A_{\parallel}|\simeq |ck_{\parallel}k_{\theta}\delta\phi/\omega|$, the threshold condition for the nonlinear process   in the $|b_L|\ll1$ limit can be estimated as:
\begin{eqnarray}
\left(\frac{\delta B_r}{B_0}\right)^2\sim \frac{\gamma_L\gamma_G}{\omega^2_0}\frac{k^2_{\parallel,0}}{k^2_L}\frac{4}{\epsilon_0}\sim O(10^{-9}).
\end{eqnarray}
In estimating the threshold condition,  typical tokamak plasma parameters are used. The nonlinear cross-section of the analyzed processes  are   comparable with or greater than other   wave-wave coupling channels for TAE saturation  in the short wavelength ($k_rk_{\theta}\rho^2_i>\omega/\Omega_{ci}$ ) kinetic regime \cite{LChenEPL2011}, e.g., zero frequency zonal structure generation \cite{LChenPRL2012} and ion induced scattering \cite{ZQiuNF2018}. Thus, the processes discussed in the present work are relevant and competitive for TAE nonlinear dynamics, where, for a realistic description, all the various processes must be accounted for on the same footing, as it is argued below.

\subsection{Relevant tokamak plasma parameter regimes}

Several processes with comparable scattering cross-section may contribute equally to the nonlinear saturation of TAE. As a result, the nonlinear dynamics of TAE can depend quite sensitively on the tokamak plasma parameter regimes and corresponding threshold conditions. Thus, the   parameter regimes for each process  to occur and possibly dominate  must be well understood. For the   processes discussed in this paper to take place, several conditions are required  as addressed below.

First, for resonant decay, both GAM and TAE/LKTAE should be weakly damped normal modes of the system. For GAM dominated by thermal ion transit resonance, weak ion Landau damping requires that GAM frequency be larger than thermal ion transit frequency, $\omega_G>\omega_{tr,i}$, which yields $q\sqrt{7/4+\tau}>1$, that is a usually satisfied condition. Meanwhile, for the TAE/LKTAE to be weakly damped by thermal ion Landau damping,   e.g., by ion sideband resonance with $v_{\parallel,res}=V_A/3$, for which the TAE/LKTAE damping rate is  $\gamma_T/\omega_T\propto\exp(-v^2_{\parallel,res}/v^2_{t,i})=\exp(-1/(9\beta_i))$ \cite{RBettiPoFB1992}, one reasonable upper-bound for  $\beta_i$ (e.g., $\gamma_T/\omega_T\lesssim 10^{-2}$) can thus be  $\beta_i<3\%$.

Second, for mode-mode coupling processes in the short wavelength kinetic regime to occur and dominate other mode-mode couplings in the MHD limit \cite{TSHahmPRL1995}, the condition $k_rk_{\theta}\rho^2_i>\omega/\Omega_{ci}$ is required. For TAE excited by circulating EPs, one typically has  $k_{\theta}\rho_i q\sqrt{T_E/T_i}  \simeq 1$; i.e., the poloidal wavelength is comparable to the circulating EP magnetic drift orbit width. On the other hand, for the short length scales that provide the dominant contribution of Reynolds and Maxwell stresses, one has $k_r\simeq k_{\theta}/\epsilon$. Thus, kinetic regime corresponds to $(T_i/T_E)/(q^2\epsilon)\gg \omega_0/\Omega_{ci}$, which is the case for typical burning plasma parameters.

Third, for the pump TAE to decay into a GAM and an LKTAE, as we discussed in Sec. \ref{sec:high_beta}, the GAM frequency must be larger than the difference between pump TAE frequency and lower accumulation  point frequency of toroidicity induced SAW continuous spectrum frequency gap, which we denote as $\omega_l$. Thus, $\omega_G>\omega_0-\omega_{\ell}\sim\lambda\epsilon\omega_A$, which gives $\beta_i>[\lambda\epsilon/(2q)]^2$ with $\lambda$   expressing the fraction of $\omega_0-\omega_{\ell}$ in units of the frequency gap width. Note that TAEs are typically localized within the lower half of the toroidicity induced SAW continuum frequency gap, and we have $0<\lambda<1/2$. This criterion, thus, sets the lower bound of $\beta_i$ for this   process to occur. In the opposite limit, with $\beta_i<[\lambda\epsilon/(2q)]^2$, i.e., $\omega_G<\omega_0-\omega_{\ell}$, the lower sideband is then a TAE lower sideband within the toroidicity induced SAW continuum frequency gap, as we discussed in Sec. \ref{sec:low_beta}. \footnote{Note that, in Ref. \citenum{ZQiuNF2018}, a similar analysis on $\beta_i$ is presented, for the optimal condition for ion-induced scattering to occur.}

In summary, the  parameter regime for this process to occur and possibly  dominate is, 1. $q\sqrt{7/4+\tau}>1$, 2. $(T_i/T_E)/(q^2\epsilon)\gg \omega_0/\Omega_{ci}$ and 3.   $[\lambda\epsilon/(2q)]^2<\beta_i<3\%$ for TAE to decay into a GAM and an LKTAE, or $\beta_i<\mbox{min}[3\%,[\lambda\epsilon/(2q)]^2]$ for TAE to decay into a GAM and a TAE lower sideband.  These conditions  also suggest  the proper setup of plasma parameters to verify this process in numerical simulations or experimental conditions.

\section{Consequences on plasma heating}\label{sec:heating}

The physics processes discussed above, provide  a new channel for transferring fusion alpha particle power to thermal plasmas. To be more specific, considering that TAE pump is resonantly excited by EPs, the ion Landau damping of the nonlinearly driven GAM, will nonlinearly transfer fusion alpha power   to thermal ions, providing  a novel ``alpha-channeling" mechanism \cite{NFischPRL1992,NFischNF1994}. On the other hand, the trapped electron collisional (or Landau) damping of the high frequency daughter wave, leads to thermal electron heating, i.e.,  the fusion alpha particle    anomalous slowing down.  To quantitatively estimate the thermal plasma heating rate, the nonlinear saturation level of GAM and the high frequency daughter wave are needed, which can be derived from equations (\ref{eq:GAM_global}) and (\ref{eq:KTAE_global}), with an additional equation for the feedbacks of the two daughter waves to the pump TAE. This aspect is neglected in Sec. \ref{sec:nl_dr} focusing on the exponential  growth stage of the parametric decay process. The pump TAE equation can be derived closely following equation (\ref{eq:KTAE_global}):
\begin{eqnarray}
\hat{\mathscr{E}}_0A_0=-i(c/B_0)k_{0,\theta}\hat{\alpha}_0A_GA_h,\label{eq:pumpTAE_eigenmode}
\end{eqnarray}
with
\begin{eqnarray}
\hat{\mathscr{E}}_0\equiv \int dr |\Phi_0|^2 \mathscr{E}_0\nonumber
\end{eqnarray}
being the eigenmode dispersion function of pump TAE, and
\begin{eqnarray}
\hat{\alpha}_0\equiv \int dr |\Phi_0|^2|\Phi_h|^2 k_G \left[\frac{\Gamma_h-\Gamma_G}{\omega_0}-\frac{(1-\Gamma_0)\sigma_h}{\sigma_0\omega_0}\right].\nonumber
\end{eqnarray}

In the local limit, the three-wave nonlinear  envelope  equations  can then be  derived by expanding equations (\ref{eq:GAM_global}), (\ref{eq:KTAE_global}) and (\ref{eq:pumpTAE_eigenmode}) along their respective characteristics, and be  cast as
\begin{eqnarray}
\left(\partial_t-\gamma_0\right)A_0&=&-\frac{c}{B_0\partial_{\omega_0} \hat{\mathscr{E}}_{0,R}}k_{0,\theta}\hat{\alpha}_0A_GA_h,\label{eq:TAE_dynamic}\\
\left(\partial_t+\gamma_{G^*}\right)A_G&=&-\frac{c}{2B_0b_G}k_{0,\theta}\hat{\alpha}_GA_{0^*}A_h, \label{eq:GAM_dynamic}\\
\left(\partial_t+\gamma_h\right)A_h&=& \frac{c}{B_0\partial_{\omega_h}\hat{\mathscr{E}}_{h,R}}k_{0,\theta}\hat{\alpha}_hA_{G^*}A_0.\label{eq:LKTAE_dynamic}
\end{eqnarray}
Here, $\gamma_0$ is the linear growth rate of pump TAE due to, e.g., resonant EP drive.   Equations (\ref{eq:GAM_dynamic}) and (\ref{eq:LKTAE_dynamic}) are used in Sec. \ref{sec:nl_dr} for deriving the parametric dispersion relation, equation (\ref{eq:parametric_DR}), letting $\partial_t=\gamma$.  The pump TAE dynamic equation (\ref{eq:TAE_dynamic}), with  an interesting one-to-one correspondence to equation (\ref{eq:LKTAE_dynamic}), has a negative sign on the right hand side unlike equation (\ref{eq:LKTAE_dynamic}), showing energy conservation in the three-wave coupling system.

The saturation levels of  high frequency daughter wave and GAM can be estimated from the fixed point solution of the above coupled  equations. Note that  this does not mean that the coupled three equations will necessarily exhibit fixed point solutions. In fact,
the nonlinear evolution of the driven-dissipative system, may be characterized by rich dynamics such as limit-cycle oscillations, period-doubling and route to chaos \cite{DRussellPRL1980}.  By taking $\partial_t=0$, the high frequency daughter wave and GAM saturation level can be derived from the  fixed point solutions as
\begin{eqnarray}
|A_h|^2&=& -2\gamma_0\gamma_Gb_G\partial_{\omega_0}\hat{\mathscr{E}}_{0,R}/((c/B_0)^2k^2_{0,\theta}\hat{\alpha}_0\hat{\alpha}_G),\nonumber\\
|A_G|^2&=& \gamma_0\gamma_h\partial_{\omega_h}\hat{\mathscr{E}}_{h,R} \partial_{\omega_0}\hat{\mathscr{E}}_{0,R}/((c/B_0)^2k^2_{0,\theta}\hat{\alpha}_0\hat{\alpha}_h).\nonumber
\end{eqnarray}

The corresponding thermal ion and thermal electron heating rate  can be derived, and yield
\begin{eqnarray}
P_i&=&\frac{1}{\hat{\alpha}_h}\frac{n_0e^2}{T_i}\frac{\gamma_0\gamma_G\gamma_h b_G\partial_{\omega_h}\hat{\mathscr{E}}_{h,R}\partial_{\omega_0}\hat{\mathscr{E}}_{0,R}}{4\pi(c/B_0)^2k^2_{0,\theta}\hat{\alpha}_0},\label{eq:ion_heating}\\
P_e&=&\frac{\omega_h}{\hat{\alpha}_G}\frac{n_0e^2}{T_i}\frac{\gamma_0\gamma_G\gamma_h b_G\partial_{\omega_h}\hat{\mathscr{E}}_{h,R}\partial_{\omega_0}\hat{\mathscr{E}}_{0,R}}{4\pi(c/B_0)^2k^2_{0,\theta}\hat{\alpha}_0}.\label{eq:electron_heating}
\end{eqnarray}
 One can then roughly estimate, the  fuel ions heating rate is, thus, of $O(\epsilon_0)$ weaker than that of  electrons.

\section{Conclusions and Discussions}\label{sec:summary}

In conclusion,  TAE decaying into a GAM and a high frequency daughter wave with the same poloidal/toroidal mode number of the pump TAE,  is investigated  as a potential mechanism for the nonlinear saturation of TAE. This channel is possible when both GAM and TAE/LKTAE are weakly damped due to  ion  Landau damping. Another key parameter in determining the TAE nonlinear evolution is $4 q^2\beta_i/\epsilon^2_0$, determining the high frequency daughter wave to be a TAE sideband within the toroidicity induced SAW continuum frequency gap \cite{ZQiuEPL2013}, or an LKTAE in the SAW continuous frequency spectrum \cite{ZQiuPRL2018}. For the TAE decay processes in the kinetic regime to dominate over mechanism in the MHD limit, $k_rk_{\theta}\rho^2_i\gg\omega/\Omega_{ci}$ is also required, which is the typical parameter for burning plasmas.

The nonlinear dispersion relation for the decay instability  is derived, which is valid for arbitrary wavelength. The conditions for the decay instability to take place,   and the threshold condition on pump TAE amplitude to overcome GAM and the high frequency daughter wave damping, are derived for the low- and high-$\beta_i$ limits, respectively.  It is found that, in the low-$\beta_i$ limit \cite{ZQiuEPL2013}, the spontaneous decay requires the pump TAE to lie within the upper half of the toroidicity induced SAW continuous frequency spectrum gap. While in the high-$\beta_i$ limit,  the condition for spontaneous decay is complicated due to the non-trivial contribution of mode structure and toroidicity. A rough estimation is made in the long wavelength limit, and we found that, this proposed channel is indeed relevant and competitive in burning plasma relevant parameter regime.   As a final remark, as the finite frequency ZS, GAM may interact with other turbulences in the plasma, e.g., drift wave turbulence, and leads to cross-scale couplings and possibly, improved confinement. This is also open for investigation \cite{ABiancalaniIAEAFEC2018}.

\section*{Acknowledgements}

This work is supported by   the National Key R\&D Program of China  under Grant No.  2017YFE0301900,
and the National Science Foundation of China under grant Nos.  11575157 and 11875233. LC acknowledges the support of US DoE grant.
This work was also carried out within the framework of the EUROfusion
Consortium and received funding from the EURATOM research and training programme
2014 - 2018 under Grant Agreement No. 633053 (Project Nos. WP15-ER/ENEA-03 and
WP17-ER/MPG-01). The views and opinions expressed herein do not necessarily reflect
those of the European Commission.

\end{document}